\documentclass[aps,prl,twocolumn,superscriptaddress]{revtex4-1}

\usepackage[utf8]{inputenc}
\usepackage[OT4]{fontenc}
\usepackage{bbm}
\usepackage{placeins}
\usepackage[dvipsnames]{xcolor}
\usepackage{graphicx}
\usepackage{hyperref}
\usepackage[export]{adjustbox}
\usepackage{mathtools}
\usepackage{amsmath}
\usepackage{amssymb}
\usepackage{physics}
\usepackage{braket}

\usepackage[colorinlistoftodos]{todonotes}
\newcounter{todocounter}

\hypersetup{pdfborder = {0 0 0}}
\hypersetup{colorlinks,citecolor=blue,filecolor=blue,linkcolor=blue,urlcolor=blue}

\graphicspath{{./Plots/}}

\begin{document}
\title{Output statistics of quantum annealers with disorder}

\author{Jonathan Brugger}
\affiliation{Physikalisches Institut, Albert-Ludwigs-Universität Freiburg,\\ Herrmann-Herder-Straße 3, D-79104 Freiburg, Federal Republic of Germany}

\author{Christian Seidel}
\affiliation{Data:Lab,Volkswagen Group, Ungererstr. 69, 80805 München, Federal Republic of Germany}

\author{Michael Streif}
\affiliation{Physikalisches Institut, Albert-Ludwigs-Universität Freiburg,\\ Herrmann-Herder-Straße 3, D-79104 Freiburg, Federal Republic of Germany}
\affiliation{Data:Lab,Volkswagen Group, Ungererstr. 69, 80805 München, Federal Republic of Germany}

\author{Filip A. Wudarski}
\affiliation{Physikalisches Institut, Albert-Ludwigs-Universität Freiburg,\\ Herrmann-Herder-Straße 3, D-79104 Freiburg, Federal Republic of Germany}
\affiliation{Institute of Physics, Faculty of Physics, Astronomy and Informatics, Nicolaus Copernicus University, Grudzi\k{a}dzka 5/7, 87–100 Toru\'n, Poland}

\author{Christoph Dittel}
\affiliation{Physikalisches Institut, Albert-Ludwigs-Universität Freiburg,\\ Herrmann-Herder-Straße 3, D-79104 Freiburg, Federal Republic of Germany}
\affiliation{EUCOR Centre for Quantum Science and Quantum Computing, Albert-Ludwigs-Universität Freiburg,\\ Herrmann-Herder-Straße 3, D-79104 Freiburg, Federal Republic of Germany}

\author{Andreas Buchleitner}
\affiliation{Physikalisches Institut, Albert-Ludwigs-Universität Freiburg,\\ Herrmann-Herder-Straße 3, D-79104 Freiburg, Federal Republic of Germany}
\affiliation{EUCOR Centre for Quantum Science and Quantum Computing, Albert-Ludwigs-Universität Freiburg,\\ Herrmann-Herder-Straße 3, D-79104 Freiburg, Federal Republic of Germany}

\begin{abstract}
We demonstrate that the output statistics of a standard quantum annealing protocol run on D-Wave 2000Q can be mimicked by static disorder garnishing an otherwise ideal device hardware -- for a 10-qubit toy Hamiltonian as well as for problem instances with thousands of qubits. A Boltzmann-like distribution over distinct output states is shown to emerge with increasing disorder strength.
\end{abstract}
\maketitle

{\it Introduction --}
Moderate size quantum annealing devices, such as D-Wave 2000Q with more than 2000 qubits, are attracting growing attention \cite{rev, mcgeoch} as the potentially first real-world implementations of the quantum computing paradigm with useful applications. Quantum annealers are designed to solve discrete optimization problems \cite{opt1, opt2} with a broad portfolio of applications ranging from traffic flow control to material design \cite{traffic, material}. They have been extensively compared to other optimization heuristics like Quantum Monte Carlo, Hamze-de Freitas-Selby, simulated annealing or other satisfiability solvers, and were claimed to outperform some of these in certain cases \cite{tunneling, localruggetness, csp}. Further suggested functionality is the efficient sampling from a Boltzmann-like distribution, which is relevant for applications of artificial neural networks in a machine learning context \cite{RBM, bayes, HM}. This feature of quantum annealing output data has so far been studied only empirically \cite{nasa1, nasa2}, without a clearly identified origin. 

Despite stunning experimental progress \cite{google}, current annealing devices are still subject to various sources of imperfections which limit possible applications -- in particular when it comes to large computational problems which constitute the relevant benchmark for any quantum computational advantage. Those error sources are essentially (i) imperfections in the initial state preparation, (ii) non-adiabatic transitions during the annealing step, (iii) a loss of coherence induced by environment coupling,  (iv) readout errors, and (v) static disorder due to limited control of the precise hardwiring of the machine \cite{Venturelli15,Zhu16,dwave}. 

Given the ever improving  degree of control on the quantum state of single qubits,  on switching profiles which warrant adiabaticity, and on the separation of coherent and incoherent time scales, one may argue that static disorder is the dominant source of errors: In particular when moving towards larger architectures with an increasing number of control parameters, it is clear that the latter can be tuned with only finite precision, such that a certain level of disorder is fundamentally  unavoidable \cite{martin-mayor15}. This is accounted for by a variety of different models using disorder [also termed disorder chaos, J-chaos, static noise, analog errors or integrated control errors (ICE)] as an essential ingredient to reproduce D-Wave output \cite{Ref3.4,Ref3.5,Ref3.6.3,Ref3.8.1,Ref3.9Add,Ref3.12}.

Our present purpose is to formulate a scalable statistical model to be directly benchmarked against D-Wave 2000Q output statistics, for variable quantum register size. While annealing into the ground state of target Hamiltonians on 10-qubit registers can still be verified by exact diagonalization on classical devices, an appropriate continuum limit formulation is adopted to handle problem instances with up to 2048 active qubits. Our model faithfully reproduces D-Wave 2000Q output statistics, with an appropriate level of static disorder as the {\em only} source of imperfections. Analysis of D-Wave 2000Q output shows that the disorder strength per qubit is constant over a broad range of register sizes, while the cumulated disorder strength increases significantly (approximately $\propto \sqrt{N}$) with the number of active qubits. As a corollary, our results establish the first quantitative explanation for near to Boltzmann distributed output, caused by a critical level of static disorder imprinted on the target Hamiltonian.

{\it Model --} 
The D-Wave machine implements the transformation of a time-dependent Hamiltonian $H(t) = A(t/\tau) H_\mathrm{in} + B(t/\tau) H_{\rm fin}$ from the initial $H_\mathrm{in} = - \sum_i \sigma_x^{(i)}$ (with Pauli matrix $\sigma_x^{(i)}$ acting on the $i$-th qubit) into the final (target, or problem) $H_\mathrm{fin}$, on a tunable time scale $\tau$, with monotone \cite{dwave} switching functions  $A(t/\tau)$ and $B(t/\tau)$ \footnote{$A(s) = -15.42s^3 + 38.33s^2 - 32.15s + 9.13$, $B(s) = 11.07s^2 + 2.19s + 0.11$, both $A$ and $B$ given in GHz. In particular, $A(0)\gg B(0)$ and $A(1)\ll B(1)$.}. The annealing protocol consists in the (ideally) adiabatic evolution of the ground state of $H_\mathrm{in}$ into that of $H_{\rm fin}$, during the (sufficiently long) annealing time $\tau$, under the action of $H(t)$. On output, the configuration of the quantum register qubits' final states -- spin up or down -- is read out and converted into the output state's energy, by evaluation of the target Hamiltonian's energy with respect to that very output state.

The Ising-type \cite{dwave} target Hamiltonian
\begin{equation}
	H_{\rm fin} =\alpha \sum_{i=1}^Nh_i \sigma_z ^{(i)} + \alpha\sum_{i,j\in \mathcal{G}} J_{ij}\sigma_z^{(i)}\otimes \sigma_z^{(j)}\label{eq1}
\end{equation}
is physically generated by switching on local control fields and non-local interaction terms, and we additionally introduce a global scaling factor $\alpha$ to control the average level spacing \cite{Ref3.6.3, Ref3.7, Ref3.12, Ref3.13}. The machine's (limited) connectivity defines the index set $\mathcal{G}$ of the second sum in \eqref{eq1}, and is restricted by the so-called {\it Chimera} graph architecture \cite{dwave}. The optimization problem to be solved is encoded \cite{QuboChimera,GraphPart,QuboTSP} in the choice of the  local energy splittings $\alpha h_i\in [-2,2]$ and coupling strengths $\alpha J_{ij}\in[-1,1]$, respectively, all in units of the final value $B(1)$ of the annealing protocol's switching function. For all D-Wave data which enter our subsequent analysis we have $B(1)=13.37\, \rm GHz$ \footnote{Private Communication by D-Wave Systems.}.
\begin{figure}
	\centering
	\includegraphics[width = 0.485\textwidth,left]{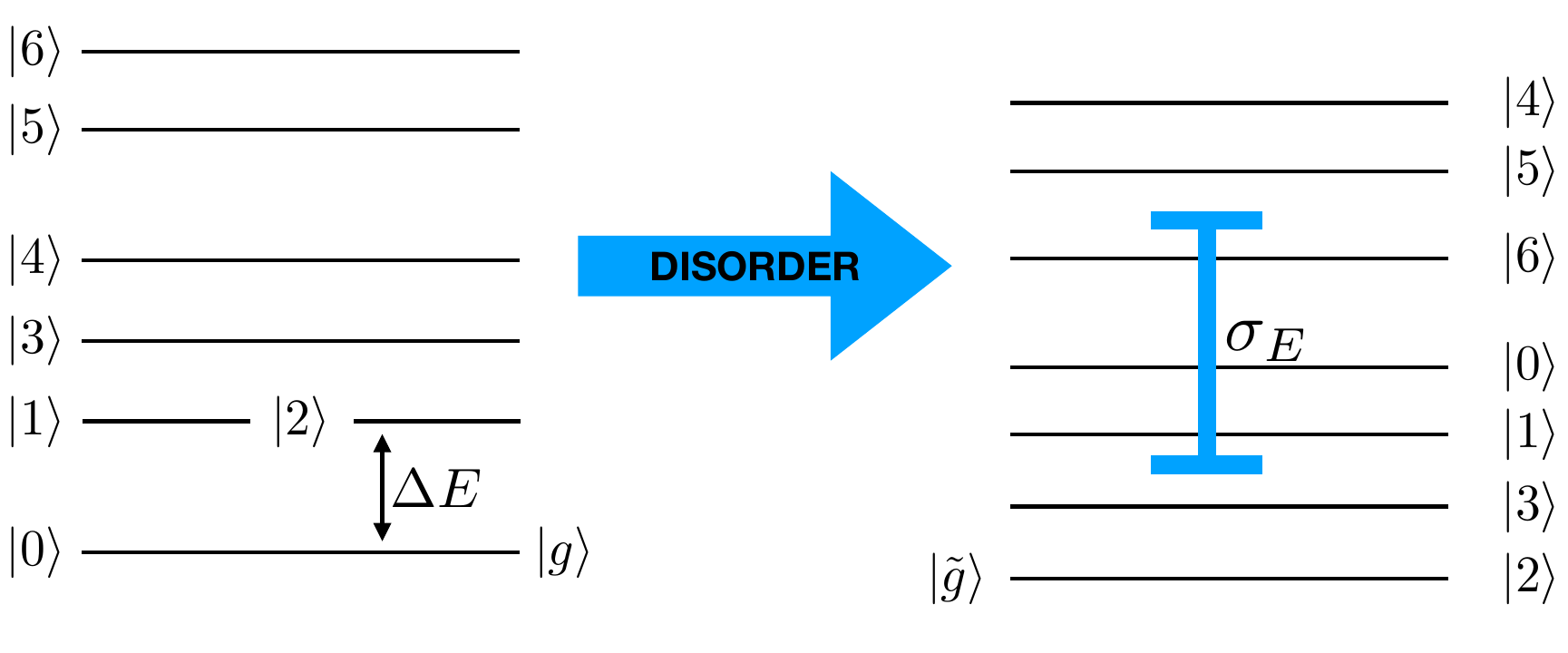}
	\caption{Sketch of the energy spectrum of the ideal (left) vs. a disordered realization of the target Hamiltonian (right): For energetic disorder strengths $\sigma_E$ similar or larger than the low energy level spacing $\Delta E$, the energetic ordering of the eigenstates of $H_{\rm fin}$ in \eqref{eq1} can change, leading to a false ground state ($\ket{\tilde g} \ne \ket{g}$) and hence solution of the optimization problem encoded in $H_{\rm fin}$.}\label{Fig1}
\end{figure}

An  ideal adiabatic protocol will deterministically output the desired target Hamiltonian's ground state, if the tuning parameters $h_i$ and $J_{ij}$ are set with infinite precision. Ubiquitous and unavoidable tuning imperfections of site energies and inter-site coupling, however, will de facto induce a {\em statistical distribution} of target Hamiltonians when sampling over a sequence of annealing runs, and, consequently, also a statistical distribution of output states and energies. The reliability of the annealing protocol's output under such conditions will essentially depend on the competition between the disorder strength and the density of states in the low-energy range, since this can induce undesired changes of the energetic ordering of different output register configurations, and thus an incorrect identification of the solution state, as illustrated in Fig.~\ref{Fig1} \cite{Venturelli15,Zhu16}. To assess the protocol's reliability, we therefore need to quantify the probability for excited states of the ideal target Hamiltonian to be transformed into the ground state \cite{Ref3.4} of its disordered realizations -- which are effectively sampled over.

Let us make these considerations more quantitative with the following model: We assume perfectly adiabatic evolution (at vanishing physical temperature, i.e. under the assumption of unitary dynamics) from the $N$-qubit ground state of $H_\mathrm{in}$ into the ground state $\ket{\tilde{g}}$ of a disordered realization $\tilde{H}_{\rm fin}$ of the ideal target $H_{\rm fin}$ with precisely specified parameters $h_i$ and $J_{ij}$. $\tilde{H}_{\rm fin}$ is different (this is the meaning of ``disorder") in each run of the annealing protocol, and constructed according to \eqref{eq1}, with energy splittings and couplings normally distributed around their ideal values, with standard deviations $\sigma_h$ and $\sigma_J$. In other words, we substitute  $\alpha h_i\to \tilde{h}_i= \alpha h_i + X_i$ and $\alpha J_{ij}\to \tilde{J}_{ij} = \alpha J_{ij}+Y_{ij}$, with $X_i\sim \mathcal{N}(0,\sigma_h)$ and $Y_{ij}\sim \mathcal{N}(0,\sigma_J)$ normally distributed around zero, ignore cross talk between (next to) adjacent qubits, as well as background susceptibility \cite{dwave}, and optimize $\sigma_h$ and $\sigma_J$ with respect to D-Wave output (independent of the actual values of $h_i, J_{ij}$ \cite{Ref3.4,Ref3.12}) around those values $\sigma_{J,0}=0.035$ and $\sigma_{h,0}=0.05$ [again in units of $B(1)$] reported for a specific experimental realization of D-Wave 2 in \cite{errors,alejandro}. Disorder thus induces a finite standard deviation of the target Hamiltonian's energies, given by
\begin{equation}
	\sigma_E = \gamma \, \sigma_{E,0} = \gamma \, \sqrt{\sigma_{h,0}^2 N+\sigma_{J,0}^2|\mathcal{G}|}\label{eq2}
\end{equation}
(see Fig.~\ref{Fig1}), with $N$ and $|\mathcal{G}|$ the number of qubits and active connections, respectively, and $\gamma = \sigma_E / \sigma_{E,0} = \sigma_h / \sigma_{h,0} = \sigma_J / \sigma_{J,0}$ a disorder scaling factor with respect to $\sigma_{E,0}$ as deduced from $\sigma_{J,0}$ and $\sigma_{h,0}$. Since both $\tilde{H}_{\rm fin}$ and $H_{\rm fin}$ are diagonal in the $\sigma_z$-basis, the identification of the ground state $\ket{\tilde{g}_i}$ of the $i$-th random realization of $\tilde{H}_{\rm fin}$ gives rise to a statistical sample of output energies $E_i = \bra{\tilde{g}_i} H_{\rm fin} \ket{ \tilde{g}_i }$.

\begin{figure}
	\centering
	\includegraphics[width = 0.47\textwidth,left]{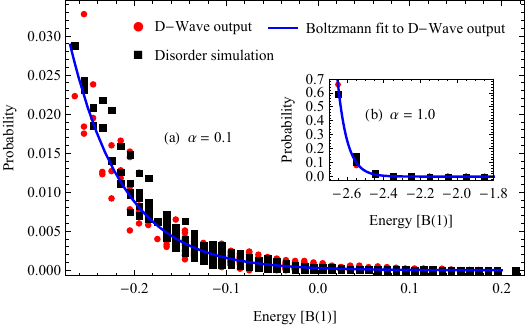}
	\caption{(Color online) Comparison of the D-Wave 2000Q output distribution ($10\,000$ annealing runs per parameter set) to the predictions of our disorder model ($100\,000$ disorder realizations per parameter set), for the longest annealing time $\tau = 2000\, \mu s$:	Possible output energies are the eigenvalues of the ideal target Hamiltonian $H_{\rm fin}$ in \eqref{eq1}. For disorder strengths $\sigma_h = 0.034$ and $\sigma_J = 0.024$, very good agreement of D-Wave and model data is observed, for any energy scale $\alpha$. This manifests in small values of the Jensen-Shannon Divergence (JSD), ranging from $0.02\,\%$ to $4.27\,\%$ for $\alpha \in \{0.1, 0.2, 0.5, 1.0, 2.0, 4.0\}$ (only $\alpha \in \{0.1, 1.0\}$ depicted here, with JSD $4.27\,\%$ and $1.28\,\%$, respectively) \cite{JBMSc}. For small $\alpha \leq 1.0$, Boltzmann-like spreading of the output data over a broad range of ground state energies is apparent, and allows to extract an {\em effective} inverse temperature $\beta =1/k_{\rm B}T_{\rm eff}$, as shown by the blue lines in both plots [fitted against D-Wave output, yielding $\beta = 17.1\ (18.8)$ for $\alpha = 0.1\ (1.0)$. For each eigenvalue of the ideal Hamiltonian, there are in general several data points, due to the degeneracy of the associated eigenspaces. This guarantees the normalization of the distributions. $\sigma$ and $\beta$ are given in units of the switching function's final value $B(1)=13.37\, \rm GHz$ and of its inverse, respectively.}\label{Fig2}
\end{figure}

{\it Results (small systems) --} We now scrutinize this model using a ten qubit target Hamiltonian $H_{\rm fin}$ \footnote{We use the qubits $\{0, 1, 2, 3, 5, 8, 9, 10, 12, 13\}$, with $h_0 = -0.25$, $h_1 = -0.1$, $h_2 = -0.25$, $h_3 = -0.1$, $h_5 =  0.25$, $h_8 = -0.25$, $h_9 = -0.1$, $h_{10} = -0.25$, $h_{12} = 0.0$, $h_{13} = 0.25$ and $J_{0,4} = -0.1$, $J_{1,4} = -0.1$, $J_{2,4} = -0.1$, $J_{0,5} = 0.2$, $J_{1,5} = -0.25$, $J_{2,5} = 0.2$, $J_{4,12} = -0.1$, $J_{8,12} = 0.2$, $J_{9,12} = -0.1$, $J_{10,12} =  -0.25$, $J_{5,12} = 0.1$, $J_{8,13} = 0.1$, $J_{9,13} = -0.25$, $J_{10,13} = 0.2$.}. On a classical computer, we extract output energies $E_i$ as described above, for $100\,000$ disordered realizations $\tilde{H}_{\rm fin}$ of $H_{\rm fin}$ per parameter set $(\sigma_h, \sigma_J, \alpha)$. This provides the probability for the different eigenstates of $H_{\rm fin}$ (identified by their unperturbed energies) to be transformed into the ground state of the disordered Hamiltonian $\tilde{H}_{\rm fin}$, and, hence, to lead to a false result of the annealing protocol. Note that these probabilities are solely given by the distribution of ground states, defined by $H_{\rm fin}$ and $\sigma_{h/J}$ alone. In particular, we do not simulate the unitary dynamics of the system, but rather assume -- aside from disorder -- ideal performance of the D-Wave hardware in correctly identifying the ground state of the disordered realization $\tilde{H}_{\rm fin}$. To validate our hypothesis that disorder alone suffices to reproduce the salient features of the D-Wave output, we quantify the agreement of the above simulation data with the latter. D-Wave data are generated for identical $H_{\rm fin}$, with tunable energy scales $\alpha \in \{0.1, 0.2, 0.5, 1.0, 2.0, 4.0\}$, annealing times $\tau/\mu s \in \{1,\ 10,\ 20,\ 50,\ 100,\ 500,\ 1000,\ 1500,\ 2000 \}$, and $10\,000$ annealing runs per parameter set. Since we want to distill the impact of static disorder from the output data, available post-processing settings were turned off \cite{dwave2}, and we also disregard existing error correction and mitigation schemes \cite{Ref3.5, Ref3.7, Ref3.12, Ref3.13} which can attenuate the practical consequences of disorder.

To benchmark our model against D-Wave data, we use the Jensen-Shannon divergence \cite{JSD1,JSD2,KLD} which measures the squared distance $\mathrm{JSD}(P||Q) \in [0,1]$ of two distributions $P$ and $Q$ \footnote{The Jensen-Shannon divergence is defined as $\mathrm{JSD}(P||Q) = \mathrm{JSD}(Q||P) = \left[ D(P||M) + D(Q||M) \right] / 2$, with $M = (P + Q) / 2$ and the Kullback-Leibler divergence $D(P||Q) = \sum_{i \in I} P(x_i) \log\left[ P(x_i) / Q(x_i) \right]$ for probability distributions $P$ and $Q$ on a probability space $X = \{x_i|i \in I\}$.}. The smaller JSD, the closer the distributions; $\mathrm{JSD}(P||Q) =0$ certifies the identity of $P$ and $Q$. Figure \ref{Fig2} illustrates this comparison for the longest annealing time $\tau = 2000 \, \rm \mu s$, and for two exemplary energy scales $\alpha$. For each value of $\alpha$, we performed numerical simulations with disorder strengths $\sigma_{h/J} = \gamma \, \sigma_{h/J,0}$ optimized (by minimizing the JSD of the D-Wave 2000Q output with respect to our simulation) around the literature values \cite{errors,alejandro}, to probe the sensitivity of the comparison on these important model ingredients -- which, however, are not quantified in the accessible D-Wave 2000Q documentation \cite{dwave}. We find best agreement for $\gamma = 0.677$, suggesting slightly weaker disorder than previously reported.

Our disorder model fits the D-Wave output distribution rather well, as quantified by JSD values ranging from $0.02\,\%$ to $4.27\,\%$ [we attribute remaining discrepancies to residual effects not accounted for in our model, such as items (i)-(iv) in the Introduction above]. A comparison of the results for different annealing times \cite{JBMSc} suggests that for reasonably large $\tau \gtrsim 100 \, \mu s$, the evolution is essentially adiabatic, since further increase of $\tau$ does not change the output distributions. For very small $\tau$, however, we observe a significant broadening of the output distributions, indicative of non-adiabatic transitions on top of the disorder-induced reshuffling of output states \cite{JBMSc}. Finally, we observe that, except for too large an energy scale $\alpha$ (i.e., disorder strengths negligible with respect to $\Delta E$, see FIG.~\ref{Fig1}), the output statistics is well approximated by a Boltzmann distribution  
\begin{equation}
	p_k = \frac{\exp(-\beta E_k)}{\sum_i \exp(-\beta E_i)}\,,\label{eq3}
\end{equation}
with $p_k$ the probability to measure an output state with energy $E_k$, and $\beta=1/k_BT_{\rm eff}$, with Boltzmann constant $k_B$ and finite \emph{effective} temperature $T_{\rm eff}$. We conclude that -- for a small quantum register with ten active qubits -- the overwhelming contribution to the Boltzmann-like output stems from a finite disorder strength, amended by mild non-adiabatic contributions for the very shortest annealing times.

\begin{figure}
	\centering
	\includegraphics[width = 0.47\textwidth,left]{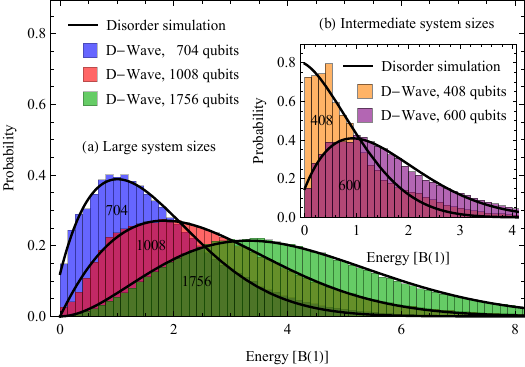}
	\caption{(Color online) Output energy distributions (histograms) of D-Wave 2000Q (50 random problem Hamiltonians $H^{(n)}_{\rm fin}$ per system size $N$, $10\,000$ annealing runs per problem Hamiltonian, annealing time $\tau = 2000\, \mu s$) compared to the continuous version $P(E)$ of our disorder model (continuous lines): D-Wave output is sampled from all target Hamiltonians $H^{(n)}_{\rm fin}$ per system size, after	aligning their individual output energy spectra to ground state	energy zero. The disorder model $P(E)$ is fitted, with excellent agreement (quantified in Fig.~\ref{Fig4}), against D-Wave output by suitable tuning of the parameters $\{a_j\}$ and $\sigma_E$. All energies are given in units of $B(1) = 13.37\, \rm GHz$.}\label{Fig3}
\end{figure}
\begin{figure}[t!]
	\centering
	\includegraphics[width = 0.475\textwidth,center]{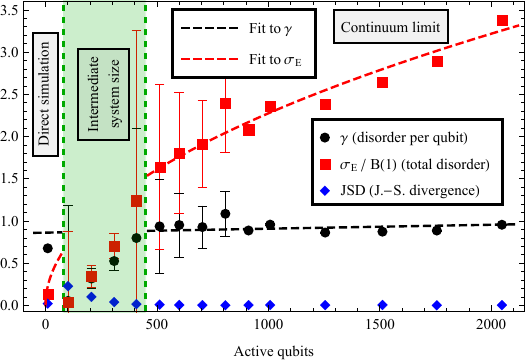}
	\caption{(Color online) Energetic disorder strength $\sigma_E$, disorder strength $\gamma$ per qubit, and JSD (Jensen-Shannon divergence) distance between D-Wave output distribution and the best fit by our model, vs.	register size $N$: For large $N$, $\gamma$ remains essentially constant, as indicated by the linear fit, and close to unity -- the previously reported disorder level \cite{errors,alejandro} [see \eqref{eq2}]. $\sigma_E$ systematically increases with system size $\propto\sqrt{N}$, due to the accumulation of the individual qubits' errors, by \eqref{eq2}. Results for $N = 10$ and  $N \geq 101$	are extracted from Figs.~\ref{Fig2} and~\ref{Fig3}, respectively. Note that, strictly speaking, the continuum model is inapplicable for $N \lesssim 500$, as indicated by increasing JSD	for smaller $N$, and therefore only data for $N \geq 500$ and $N = 10$ were used for the fits. For $N \gtrsim 500$ we obtain excellent agreement with D-Wave 2000Q output, quantified by JSD values below $1\,\%$. Error bars are deduced from the fits in Fig.~\ref{Fig3} and, for large $N$, of vanishing size.}\label{Fig4}
\end{figure}

{\it Results (large systems) --}
After this first consistency check of our disorder model against real D-Wave 2000Q data, we move on to larger register sizes, with $N \in \{$101, 206, 308, 408, 512, 600, 704, 807, 911, 1008, 1255, 1512, 1756, 2048$\}$ active qubits. Exhaustive sampling of D-Wave output or on a classical computer now is prohibitive, due to the exponentially increasing number $2^N$ of eigenstates, and concomitant, exponentially decreasing hitting probabilities for individual eigenstates on output. To statistically characterize the impact of a finite disorder strength on D-Wave 2000Q annealing performance, we therefore adopt a slightly different strategy: For the generation of statistically robust samples of D-Wave data, we create 50 (rather than one single, for $N=10$ above), randomly defined target Hamiltonians $H_{\rm fin}^{(n)}$ of type (\ref{eq1}) per system size, with $h_i \in [-2,2]$ and $J_{ij} \in [-1,1]$ uniformly distributed for all active qubits and associated couplings. Each of these altogether 700 target Hamiltonians is then programmed on D-Wave 2000Q, and output distributions are sampled through $10\,000$ annealing runs per $H_{\rm fin}^{(n)}$, with $\tau = 2000 \, \mu s$. Upon aligning each individual target Hamiltonian's output spectrum to the same minimal output energy zero, we generate a single, normalized output energy distribution $P(E)$ from the collective sample of 500\,000 output energies per register size. According to the same considerations as for small $N$ above, this distribution must be controlled by the competition between the energetic disorder strength $\sigma_E$ of the D-Wave hardware and the spectral density.

Assuming uncorrelated \cite{JBMSc} (disorder-induced) fluctuations of the individual eigenenergies of the disordered realizations $\tilde{H}_{\rm fin}^{(n)}$ of the randomly chosen target Hamiltonians, it can be shown \cite{JBMSc} that, in the limit of a quasi-continuous energy spectrum, $\sigma_E$ is the sole parameter to determine the probability
\begin{equation}
		p(E) = - C_N \log\left[ \frac{1 + \erf\left( \frac{E}{\sqrt{2} \sigma_E} \right)}{2} \right] \label{eq4}
\end{equation}
($\erf$ the error function and $C_N$ a normalization constant) for a target Hamiltonian's eigenstate with energy $E$ to be turned, by disorder, into the ground state. The output energy probability distribution $P(E)= p(E)d(E)$ generated on D-Wave 2000Q for given $N$ then follows by multiplication of $p(E)$ with the density of states $d(E)$, where we choose, in the presently relevant low energy range, the ansatz $d(E) = a_0 + a_1 \, E + a_2 \, E^2$.

Figure \ref{Fig3} nicely confirms the reproducibility of D-Wave 2000Q output by  our model (where $\{a_j\}$ and $\sigma_E$ are fitting parameters), for variable $N$. For system sizes of $N \gtrsim 500$ we obtain excellent agreement, quantified by JSD values below $1 \, \%$. For smaller $N$, the description in terms of a continuous density of states is no longer valid, as evidenced by larger JSD values, as well as by hitting probabilities above $\gtrsim 10 \, \%$ for single output states. The best-fitting values \footnote{For system sizes $500 \leq N \leq 800$ the fit quality is -- despite excellent agreement in terms of JSD values -- relatively insensitive to the disorder strength $\sigma_E$, resulting in large error bars, which, however, rapidly vanish for larger systems.} for $\sigma_E$ quantify the total amount of disorder in the D-Wave device, for different $N$, see Fig.~\ref{Fig4}. As to be expected from \eqref{eq2}, $\sigma_E\propto\sqrt{N}$ for large $N$, while the disorder $\gamma$ per qubit remains essentially constant, independently of $N$, with a value very close to previously reported results \cite{errors,alejandro}.

The quantitative agreement between D-Wave 2000Q data and our (discrete and continuous) disorder models, for register sizes from $N = 10$ to 2048, demonstrates that the overwhelming contribution to the statistical scatter can be explained by finite disorder $\gamma$ per qubit, which is here directly extracted from the statistics' scaling behavior with $N$. Since our results are consistent with the assumption of adiabatic and unitary annealing dynamics (properties which, however, are not scrutinized by our approach), this suggests that corrections due to other sources of imperfections [see, e.g., (i-iv) above] are essentially negligible. Finally, the probability density $p(E)$ in \eqref{eq4} -- uniquely characterized by the total disorder strength $\sigma_E$ -- is well approximated by a Boltzmann distribution \eqref{eq3}, and thereby offers a quantitative explanation for earlier observations \cite{nasa1, nasa2}.

{\it Acknowledgments --} J. B. thanks the Studienstiftung des deutschen Volkes for support; F. W. is indebted to the Polish Ministry of Science and Higher Education program ``Mobility Plus'' (Grant No. 1278/MOB/IV/2015/0); C. D. acknowledges the Georg H. Endress foundation for financial support. Research was co-funded by Volkswagen Group, department Group IT. Access to D-Wave output data became possible through the Volkswagen Group.



%

\end{document}